\documentclass[reprint,amsmath,amssymb,aps,prl,twocolumn]{revtex4-2}
\usepackage{graphicx}
\usepackage{dcolumn}
\usepackage{bm}
\usepackage[colorlinks=true, allcolors=blue]{hyperref}
\usepackage{mathrsfs}%
\usepackage[title]{appendix}%
\usepackage{xcolor}%
\usepackage{textcomp}%
\usepackage{booktabs}%
\usepackage{algorithm}%
\usepackage{algorithmicx}%
\usepackage{algpseudocode}%
\usepackage{listings}%
\usepackage{float}
\usepackage{siunitx}
\usepackage{microtype} 

\begin{document}

\title{Electron Rephasing in a Truncated Tunable Plasma Channel}

\author{Arujash Mohanty}
\thanks{These authors contributed equally; Corresponding email: \href{mailto:arujash.mohanty@weizmann.ac.il}{arujash.mohanty@weizmann.ac.il}}

\author{Santhosh Krishnamurthy}
\thanks{These authors contributed equally; Corresponding email: \href{mailto:arujash.mohanty@weizmann.ac.il}{arujash.mohanty@weizmann.ac.il}}

\author{Yinren Shou}
\email{shouyinren@fudan.edu.cn}

\author{Sheroy Tata}

\author{Anton Golovanov}

\author{Anda-Maria Talposi} 

\author{Aaron Liberman}

\author{Eyal Kroupp}

\author{Victor Malka}
\email{victor.malka@weizmann.ac.il}

\affiliation{%
 Department of Physics of Complex Systems, Weizmann Institute of Science, Rehovot 7610001, Israel
}%

\date{\today}

\begin{abstract}

We report the acceleration of stable electron beams with energies up to 1.5 GeV using a 50 TW laser pulse guided in a preformed plasma channel generated by an axiparabola-focused heater pulse. Varying the channel length changes the plasma density beyond the channel end and produces distinct electron spectra. A short channel causes a large decrease in bubble size after the truncated channel, leaving only part of the electron bunch in the accelerating phase and resulting in a quasi-monoenergetic peak. For a longer channel ending near the target density down-ramp, the smaller density increase rephases and further accelerates a larger fraction of the bunch, producing a continuous high-charge spectrum. Hydrodynamic and particle-in-cell simulations reproduce the observed spectral evolution. Together, the measurements and simulations identify channel length as a key control parameter in guided laser-wakefield acceleration.

\end{abstract}

\maketitle

Laser-wakefield accelerators sustain electric fields on the order of 100 GV/m, making them a tabletop-scale alternative to conventional accelerators \cite{tajima1979laser,Malka2002science,Ma2025PhaseSpace}. Efficient wakefield acceleration requires maintaining the laser intensity and keeping the electrons in the accelerating phase of the wake. The acceleration length is therefore limited primarily by laser diffraction, pump depletion, and electron dephasing \cite{Esarey2009RMP,lu2007guiding,Streeter2022Efficiency}. Laser diffraction can be mitigated by tailoring the transverse plasma density profile to form an optical channel with an on-axis density minimum and an approximately parabolic radial profile \cite{esarey1992guiding,durfee1993guiding}. Such channels, produced by hydrodynamic expansion or capillary discharge, have enabled centimeter- to meter-scale guiding and GeV to multi-GeV electron acceleration \cite{Malka96,volfbeyn1999guiding,gaul2000production,lemos2013plasma,lopes2003dynamics,bendoyro2008plasma,capillary2000hooker,capillary2002hooker,leemans2006gev,leemans2019gev,Shalloo2018HOFI,Picksley2020ConditionedHOFI,Picksley2023TruncatedChannel,Picksley2024MatchedGuiding,Lahaye2025Waveguide,Zhu2023CurvedChannel}.

While plasma channels extend the laser propagation distance, electron dephasing is commonly controlled by tailoring the longitudinal plasma density profile. Density transitions have been used to control electron injection and beam quality in laser-wakefield acceleration (LWFA) \cite{Suk2001DensityTransition,Geddes2008DensityGradient,Schmid2010DensityTransition,Brantov2008DensityInhomogeneity,Gonsalves2011DensityTailoring,Ke2021DensityTailored,Wang2021FEL,Chang2023IntegratedLens,Steyn2026PlasmaDechirper}, and density up-ramps can rephase electrons by reducing the plasma wavelength and shifting the injected electrons toward the back of the bubble \cite{Guillaume2015Rephasing,Dopp2016EnergyBoost}. More recent studies have also demonstrated control of electron rephasing using plasma lenses, blast waves, and longitudinal density tapering with an axiparabola-based channel \cite{Gustafsson2024RephasingLens,Tsymbalov2025BlastWave,
Lahaye2026DephasingReduction}. An axiparabola is a specialized reflective optic that combines axicon-like and parabolic focusing properties to produce an extended focal line along the laser propagation axis \cite{Smartsev_2019_axiparabola,Oubrerie_2022axiparabola,Oubrerie2022GeV,Lahaye2024guiding}.

In principle, a finite-length plasma channel can simultaneously mitigate laser diffraction and control electron dephasing, since it can guide the laser within the channel while creating a longitudinal density transition at its downstream end. Nevertheless, previous studies have focused mainly on injection at the channel entrance \cite{Picksley2023TruncatedChannel} or laser evolution within the channel \cite{Picksley2024MatchedGuiding}. The role of the effective channel length, particularly where the guided channel is truncated before the end of the gas target, remains largely unexplored.

In this Letter, we report LWFA in a preformed plasma channel generated by an axiparabola-focused heater pulse. By varying the heater-beam diameter, we control the length of the axiparabola focal line and therefore the effective plasma-channel length. Changing the channel length moves the channel end along the longitudinal gas-density profile, thus changing the plasma density beyond the channel and leading to different electron-rephasing dynamics. By scanning the channel length, we identify a clear transition in the electron spectra and determine an optimal length that yields improved reproducibility and higher electron energies. The observed trend is consistent across experiments performed with both 15-mm and 10-mm gas targets. These results identify plasma-channel length as a key control parameter for electron rephasing and beam stability in guided LWFA.

\begin{figure*} 
\centering
\includegraphics[height=0.33\linewidth, width=\textwidth]{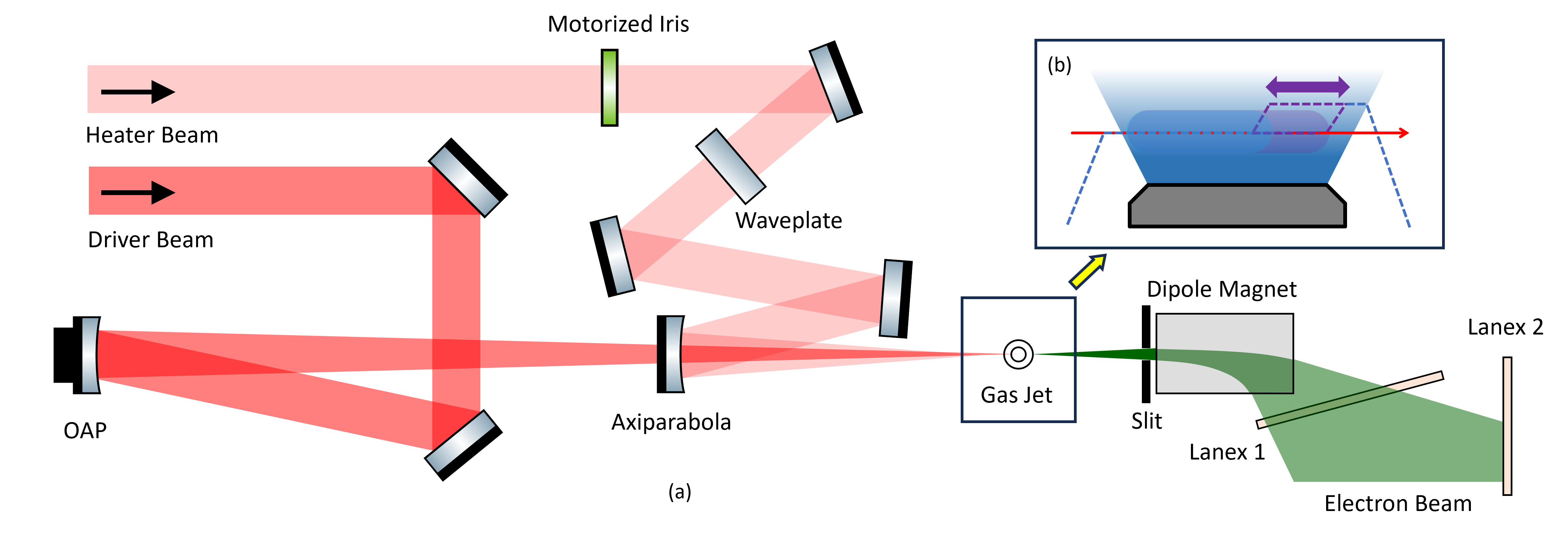}
\caption{(a) Schematic of the experimental setup. The axiparabola has a 12-mm-diameter hole on-axis through which the driver beam passes. To ensure the two beams are co-propagating along the entire nozzle length, we image the focal spot with a 5$\times$ microscope objective and a 4F imaging system onto a 16-bit CCD, and iteratively correct the pointing until the two foci coincide throughout the nozzle. (b) Side-view illustration of the gas density profiles for different channel lengths, which are controlled by varying the heater beam diameter. The red arrow indicates the laser propagation direction, while the dashed outlines illustrate the gas density along the laser axis. Varying the channel length changes the position of the channel end relative to the longitudinal gas density profile and hence the plasma density encountered by the driver beam beyond the channel.}
\label{fig:setup}
\end{figure*}

The experiment was performed on the HIGGINS $2\times100$~TW Ti:Sapphire laser system at the Weizmann Institute of Science~\cite{kroupp_2022_mre}. The experimental layout is sketched in Fig.~\ref{fig:setup}. The driver and heater pulses share a common beamline until the final amplification stage and operate at a repetition rate of \SI{1}{Hz}~\cite{kroupp_2022_mre}. Both beamlines include deformable mirrors before the compressors to correct wavefront aberrations at focus. The driver beam, which had an FWHM duration of 28\,fs and an energy of 1.4\,J after compression, was focused by a 2m focal length off-axis parabola (OAP) to a spot size of \SI{25}{\micro m}. The corresponding peak intensity was estimated to be $5.5\pm 0.5 \times 10^{18}$ $\mathrm{W/cm^2}$ equivalent to $a_0$ of 1.5 to 1.7. A co-propagating heater beam (\SI{67}{mJ}, \SI{75}{fs}) was focused by an axiparabola with a focal length of 220 mm. A quarter-wave plate was used to control polarization, which strongly affects the optical-field-ionization heating efficiency and channel geometry~\cite{lemos2013polarization}. The focal spot diameter decreased from 9 to \SI{4}{\micro m} along the focal line, with a peak intensity of approximately  $5\times 10^{15}$ $\mathrm{W/cm^2}$. The heater beam arrived at the target \SI{2}{ns} before the driver, allowing ionization and hydrodynamic expansion to form the channel. The heater and driver were focused at positions 2 and 2.5 mm, respectively, from the beginning of the gas jet, so that the driver encountered the gradual density up-ramp before entering the channel. The plasma channel length was tuned from 5 to \SI{14}{mm} by adjusting a motorized iris in the heater beam path, which changes the axiparabola focal-line length. We used conical nozzles with diameters of 15 and 10 mm (1:10 throat to exit ratio) \cite{Semushin}, puffed with 98\% He + 2\% N$_2$ as the gas target. Accelerated electrons were dispersed by a \SI{30}{cm} permanent dipole magnet with a peak field of \SI{0.97}{T}, and detected on two Lanex screens. A 5-mm slit was positioned \SI{43.2}{cm} from the gas jet to reduce the energy uncertainty from beam pointing. A complete description of the spectrometer geometry and energy reconstruction with the dual Lanexes is given in the Supplemental Material \footnote{See Supplemental Material at [URL will be inserted by publisher] for further details, which includes Refs. \cite{ulrich_Lanex}.}. 

\begin{figure*}[tb!]
\centering
\includegraphics[]{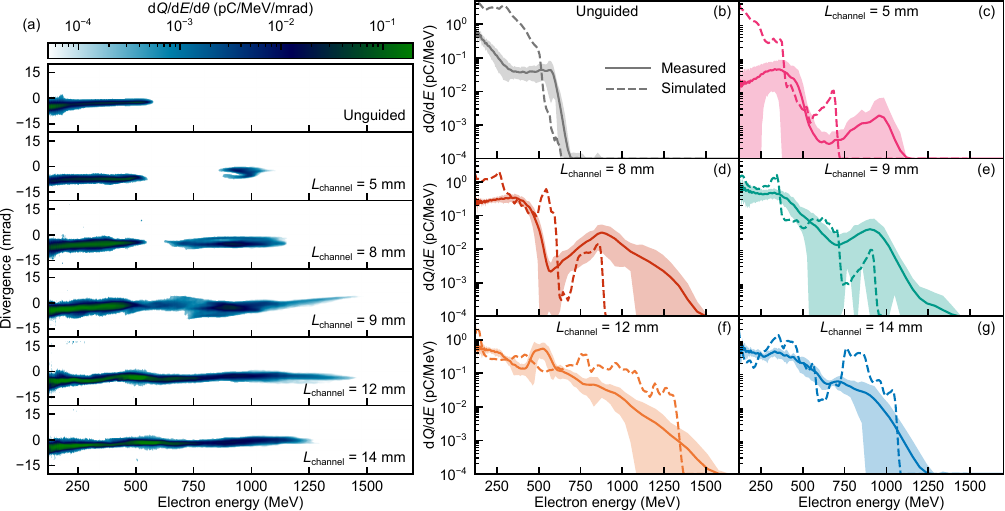}
\caption{(a) Typical electron spectra obtained with an unguided laser and with a laser guided in plasma channels of different lengths, illustrating the transition from a quasi-monoenergetic peak to continuous spectra. (b)-(g) Electron spectra averaged over 15 consecutive shots (solid curves). The shaded regions indicate one standard deviation, and the dashed curves show the simulation results.}
\label{15mm_iris_fin2}
\end{figure*}

Figure \ref{15mm_iris_fin2}(a) shows representative Lanex images of the electrons obtained by guiding the driver pulse through plasma channels with lengths of 5, 8, 9, 12, and 14 mm using a conical gas nozzle of 15 mm diameter. Over the range of electron energies shown, the angular divergence is about 2 mrad. Figures \ref{15mm_iris_fin2}(b)-(g) display the corresponding mean spectra obtained from a series of 15 consecutive shots for each channel length. All experimental data were obtained using the same main laser parameters and the same backing pressure of 21 bar, which generated a gas profile corresponding to an electron density of $4 \times 10^{18}\,\mathrm{cm^{-3}}$ at full ionization of the gas. A reference unguided case, also shown in Fig. 2, exhibited a maximum electron energy below 750\,MeV. The plasma channel increased the maximum electron energy, which reached approximately 1.5 GeV for the 12-mm channel [Fig. \ref{15mm_iris_fin2}(f)]. The best electron-beam stability and charge were also achieved for a channel length of 12 mm, with an average charge of $223 \pm 20$\,pC for electrons exceeding 120 MeV.  The same trend was observed with the 10-mm nozzle, for which the optimal channel length was 8 mm (Supplemental Material \cite{Note1}).                                                     

\begin{figure}
\centering
\includegraphics[]{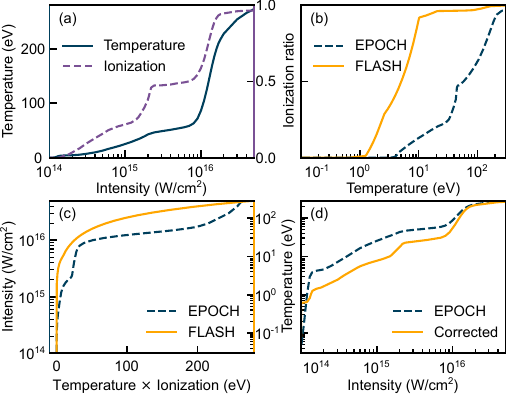}
\caption{(a) Electron temperature and ionization ratio as a function of laser intensity from 1D EPOCH simulations. (b) Relationships between ionization ratio and temperature from EPOCH and from FLASH. (c) Temperature times ionization ratio to bridge the two codes. (d) Corrected electron temperature as a function of heater intensity. The higher ionization ratio in FLASH reduces the electron temperature relative to EPOCH. Both EPOCH and FLASH simulations took into account the 2\% N$_2$.}
\label{Temp_Inten}
\end{figure}

To interpret the experimental results, we performed simulations in three stages. The laser heating was simulated using a combination of FLASH \cite{fryxell2000flash} and particle-in-cell (PIC) code EPOCH \cite{arber2015epoch}. As shown in Fig.~\ref{Temp_Inten}, we applied a correction to account for the lower ionization ratio predicted by EPOCH. The formation and subsequent 2-ns hydrodynamic expansion of the plasma channel were then simulated using FLASH. Finally, the LWFA in the resulting plasma density profile was simulated using FBPIC \cite{lehe2016fbpic}. The driver was modeled as a Gaussian pulse with an energy of \SI{1.4}{J}, an FWHM duration of \SI{28}{fs}, and a waist width of \SI{23}{\micro\meter}, corresponding to $a_0=1.7$. For each heater beam diameter, the longitudinal heater profile was reconstructed from transverse focal spot images measured at 1 mm intervals along the focal line. 

\begin{figure}[b]
\centering
\includegraphics[]{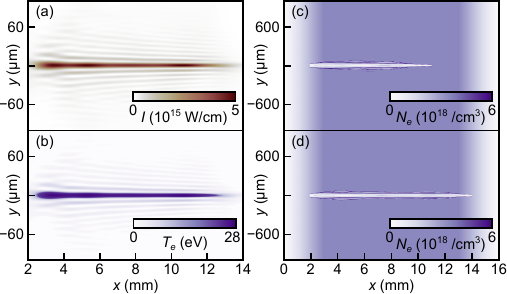}
\caption{(a) Heater laser intensity distribution and (b) electron temperature distribution after heating for a 35-mm aperture. (c) and (d) Electron density distributions at \SI{2}{ns} after the heater for a 27-mm aperture and a 35-mm aperture, corresponding to channel lengths of 8 and 12 mm, respectively.}
\label{PlasmaChannel_different_Iris}
\end{figure}

\begin{figure}[t]
\centering
\includegraphics[]{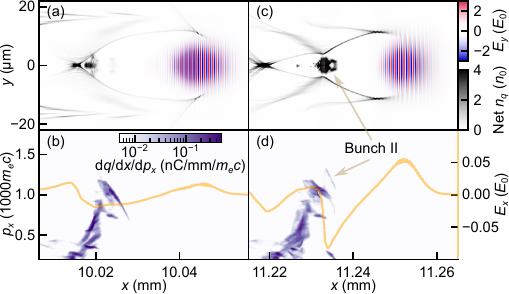}
\caption{(a) Driving laser and net charge distribution before the end of the 8-mm channel. (b) Longitudinal phase space of the accelerated electrons and the on-axis longitudinal electric field. (c), (d) Same as (a), (b) but after the end of the channel. The high plasma density beyond the channel end causes a large decrease in bubble size, leaving only part of the bunch in the accelerating phase.}
\label{PIC_iris_27}
\end{figure}

The results of the FBPIC simulations for all channel lengths are shown as dashed curves in Fig.~\ref{15mm_iris_fin2}(b)-(g). We focus on two representative cases. For heater-beam diameters of 27 and 35\,mm, the FLASH simulations give plasma channels of length $8$ and $12$\,mm, respectively, as displayed in Fig.~\ref{PlasmaChannel_different_Iris}. The quasi-monoenergetic peak observed for the 8-mm channel arises from the sharp density increase at the channel end [Fig.~\ref{PlasmaChannel_different_Iris}(c)]. The PIC simulations in Fig. \ref{PIC_iris_27} show that the bubble size decreases abruptly after $x = 10\,\mathrm{mm}$ as the driver exits the low-density channel and enters the surrounding higher-density plasma. The resulting decrease in plasma wavelength changes the phase of the accelerated electrons relative to the wake, leaving only part of the bunch in the accelerating phase of the smaller bubble. These electrons are shifted toward the back of the bubble, where they experience a stronger longitudinal accelerating field and gain additional energy before the driver diffracts, resulting in a quasi-monoenergetic peak.

\begin{figure}[ht]
\centering
\includegraphics[]{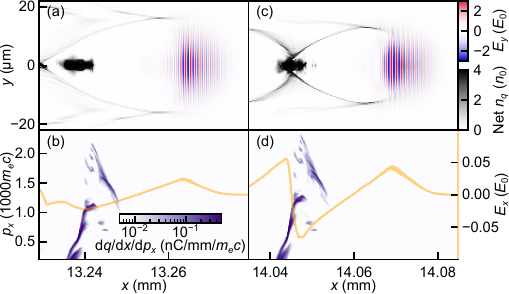}
\caption{Same as Fig. \ref{PIC_iris_27}, but for the 12-mm channel. The lower plasma density beyond the channel end leads to a more moderate decrease in bubble size, allowing a larger fraction of the bunch to be rephased and further accelerated.}
\label{PIC_iris_35}
\end{figure}

For the 12-mm channel, the plasma channel extended to approximately \(x=14\) mm, near the density down-ramp at the end of the gas target [Fig.~\ref{PlasmaChannel_different_Iris}(d)]. The plasma density beyond the channel end was therefore lower than in the 8-mm case, and the bubble size decreased more moderately. This transition rephased a larger fraction of the electron bunch into the accelerating phase, allowing the electrons to gain additional energy before leaving the plasma. Figure \ref{PIC_iris_35}(d) shows that the rephased electrons experience an accelerating field approximately twice that inside the channel [Fig.~\ref{PIC_iris_35}(b)]. The larger fraction of rephased electrons produces a continuous high-charge spectrum rather than a quasi-monoenergetic peak observed for the shorter channel. For channel lengths greater than 12 mm, the channel extended to the end of the gas target and tapered directly into vacuum. The driver therefore encountered no significant density increase before leaving the target, and the corresponding rephasing was absent or weaker, resulting in a modest reduction in electron energy.

The underlying mechanism is analogous to electron rephasing at an upward density transition, which reduces the plasma wavelength and shifts electrons toward the back of the bubble \cite{Guillaume2015Rephasing,Dopp2016EnergyBoost}. It is also more broadly related to recent works on controlling the acceleration length and dephasing through tailored plasma structures, including all-optical blast-wave control and guided density-tapered accelerators \cite{Tsymbalov2025BlastWave,Lahaye2026DephasingReduction}. In the present case, however, the rephasing section is not introduced as a separate external structure. Instead, it arises naturally when the plasma channel truncates inside the gas target and the guided driver encounters the surrounding higher-density plasma.

\begin{figure}[b]
\centering
\includegraphics{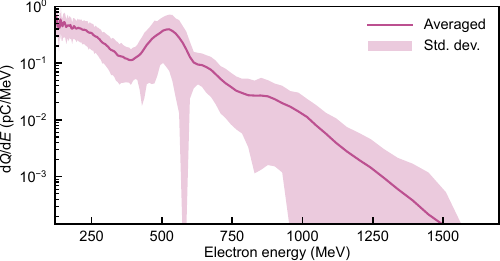}
\caption{Mean electron spectrum of 106 consecutive shots. The shaded region indicates one standard deviation.}
\label{energy_scatter_640_745_fin}
\end{figure}

To further assess the electron-beam stability at the optimal channel length, we recorded 106 consecutive shots with channel length of 12 mm using the same experimental parameters. The dual-Lanex system reduces the sensitivity of energy reconstruction to pointing variations by cross-referencing a common high-energy feature recorded on both screens. This geometry allows the electron trajectory and pointing-dependent energy to be uniquely reconstructed. Figure  \ref{energy_scatter_640_745_fin} shows the averaged electron spectra with the standard deviation of the shot series. The resultant mean electron spectrum  is continuous and has a cutoff energy of 1450 $\pm$ 42 MeV. A detailed description of the shot-series spectra and the estimation of the cutoff energy using the dual-Lanex system is provided in Supplemental Material \cite{Note1}.

We have experimentally demonstrated control of electron acceleration through rephasing at the downstream end of a preformed plasma channel. By systematically varying the plasma-channel length, we identify an optimal channel length for a given gas target length, simultaneously achieving efficient laser guiding and electron rephasing. Our results show that operation near this optimal length leads to enhanced electron energies and improved shot-to-shot reproducibility. These observations are supported by PIC and hydrodynamic simulations, which provide insight into the role of channel length in rephasing dynamics and beam evolution.
Overall, this work identifies plasma-channel length as a key control parameter in guided LWFA and provides a practical pathway toward the generation of stable, high-energy electron beams for future applications.

\begin{acknowledgments}
The research was supported by the Schwartz/Reisman Center for Intense Laser Physics, the Benoziyo Endowment Fund for the Advancement of Science, the Israel Science Foundation (contracts 2412/22 and 1267/24), and the Adelis Foundation. We would like to thank Dr. Cedric Thaury for procuring the axiparabola used in this experiment.  
\end{acknowledgments}

\noindent\textbf{Data Availability Statement.}
The data that support the findings of this Letter are not publicly available.
The data are available from the authors upon reasonable request.

\bibliography{main}

\end{document}